\documentstyle [12pt]{article}                                                  
\begin{document}                                                                
                                                                                
\baselineskip 18pt                                                              
                                                                                
\begin{center}                                                                  
{\bf
A distribution function analysis of wealth distribution } \\        
\vspace {0.3 cm}                                                                
{\it Arnab Das and  Sudhakar Yarlagadda }\\             
{Saha Institute of Nuclear Physics, Calcutta, India}             
\end{center}                                                                    
\vspace {0.5 cm}                                                                
\begin{abstract}
{We develop a general framework to analyze the distribution 
functions of wealth and income. 
Within this framework we study wealth distribution in a society
by using a model which turns on two-party trading for poor
people while for rich people interaction with wealthy
entities (huge reservoir) is relevant. At equilibrium,
 the interaction
with wealthy entities gives a power law (Pareto-like)
 behavior in the wealth distribution while the two party interaction 
gives a distribution similar to that reported earlier
\cite{chatchak,adys}.}
\end{abstract}

\vspace {1.5 cm}                                                                

{ The distribution of wealth in a nation has been a subject
of interest for a long time.
 An empirical observation due to Pareto \cite{pareto}
 states that
 the higher income group distribution has a power law tail with exponent
varying  between $2$ and $3$.
As for the  lower income group distribution it is exponential or 
 Gibb's like. The Gibb's law has been shown to be 
obtainable when trading between two people, in the absence of any savings, 
is totally random
\cite{chak,chatchak,drag,adys}. The constant
finite savings case has been
studied numerically by Chakraborti and Chakrabarti \cite{chatchak} and
analytically by us \cite{adys}. Furthermore a power law
behavior has been reported for the higher income group when
the savings are taken to be random \cite{chat,condmat}. 

{ In this paper,
we try to identify the processes that lead to the wealth distribution in
societies. Our model involves two types of trading processes - microscopic
and macroscopic. The microscopic process involves trading between
two individuals while the macroscopic one involves trading between an 
individual and the macro-system. The philosophy is that small wealth
distribution is governed by two party trading while the large wealth
distribution involves big players interacting with the macro-system.
The big players mainly interact with large entities/organizations (which
they are not directly part of) such as 
government(s), markets of other nations, etc.
 These entities/organizations are treated as making up the macro-system
in our model. 
The macro-system is thus a huge reservoir of wealth.
Whereas the poor are mainly involved in trading with other
poor individuals.
Hence, our model invokes the microscopic
channel at small wealths while at large wealths the macroscopic
channel gets turned on.

  The details of our model are as follows. An individual
possessing wealth larger than a cutoff wealth ($y_c$) trades with
a fraction [$( 1 - \lambda ) $ with $0 \leq \lambda \leq 1$] of his
wealth ($y$) with the macro-system with the latter
 putting forth an equal amount of money [$(1-\lambda) y$]. 
The trading involves the total sum $[2(1-\lambda)y]$
being randomly distributed between the individual and the macro-system.
 Thus on an average the macro-system's 
wealth is conserved.  As regards individuals that possess wealth
smaller than $y_c$ they engage in two party trading where two individuals
$i$ and $j$ put forth a fraction of their wealth  
$(1-\lambda) y_i$ and $(1-\lambda) y_j$ respectively. Then the total
money  $(1-\lambda)( y_i + y_j) $ is randomly distributed between the
two. The total money between the two is conserved in the two-party
trading process. We assume that probability of trading by
individuals having certain money is proportional
to the number of individuals with that money.

We will now develop a formalism similar to Boltzmann transport theory
so as to obtain the distribution function $f(y,\dot{y},t)$ for wealth (y), 
income ($\dot{y}$), and time (t). Similar to Boltzmann's postulate
we also postulate a dynamic law of the form
\begin{equation}
\frac{\partial f}{\partial t}
=\left \{ 
\frac{\partial f}{\partial t}
\right \}_{ext ~source} +
\left \{ 
\frac{\partial f}{\partial t} 
\right \}_{interaction} .
\label{boltz}
\end{equation}
The first term describes the evolution due to external income
sources only  and can be written as a continuity equation
\begin{equation}
\left \{ 
\frac{\partial f}{\partial t} 
\right \}_{ext~source}
=
-\frac{\partial (\dot{y} f)}{\partial y}
-\frac{\partial (\ddot{y} f)}{\partial\dot{ y}} .
\label{source}
\end{equation}

We will first analyze the microscopic channel.
As was derived earlier on by us (see Ref. \cite{adys}),
for the case involving only 
 two party interaction
with no upper cutoff,
 the  distribution function
for the money of a person  to lie between $y$ and $y+dy$ is given by
\begin{equation} 
f(y)=\int_{y}^{\infty} \frac{dL}{(1 - \lambda ) L} 
\int_{\max [0, \{y-(1-\lambda)L \} /\lambda]}^{\min [L,y/\lambda ]}
dx f(x)f(L-x) .
\label{distlamb}
\end{equation} 
We will now obtain the above equation within the Boltzmann type
transport formalism. If we assume that there is no external income
source, then
$\left \{ 
\frac{\partial f}{\partial t} 
\right \}_{ext~source} = 0$. The second (non-trivial) term
can be obtained as follows in terms of a balance equation.
\begin{equation}
\frac{\partial f}{\partial t}
=
\left \{ 
\frac{\partial f}{\partial t}
\right \}_{interaction} =gains - losses .
\label{balance}
\end{equation}
In the above equation the two terms on the right hand
side can be expressed in terms of transition rates 
$r(y_1 , y_2; y_1^{\prime}, y_2^{\prime})$ 
for a pair of persons to go from moneys $y_1 y_2$
to moneys $y_1^{\prime} y_2^{\prime}$.
Then we have on assuming that the distribution function is only
a function of wealth and time (such as when income is proportional
to one's wealth)
\begin{eqnarray}
\frac{\partial f(y_1,t)}{\partial t}
= &&
\int
r(y_1^{\prime}, y_2^{\prime};y_1 y_2)
f(y_1^{\prime},t)
f(y_2^{\prime},t)
 dy_2 dy_1^{\prime} dy_2^{\prime}
\nonumber \\
 && -
\int
r(y_1 , y_2; y_1^{\prime}, y_2^{\prime})
f(y_1,t)
f(y_2,t)
 dy_2 dy_1^{\prime} dy_2^{\prime} .
\label{inter}
\end{eqnarray}
In the above equation, conservation law requires $y_1+y_2=
y_1^{\prime}+y_2^{\prime}$. Hence in the first integral we treat $y_2$
as redundant and integrate out with respect to it to yield a
normalization constant. Similarly in the second integral
$y_2^{\prime}$ is integrated out.
Now for the transition rate in the first integral of the above equation
we have
\begin{eqnarray}
r(y_1^{\prime}, y_2^{\prime};y_1 y_2) \propto
\left \{ \begin{array}{ll}
 \frac{1}{(1-\lambda)(y_1^{\prime} +y_2^{\prime})} 
& \mbox{if $\lambda y_1^{\prime} \leq y_1 \leq 
 y_1^{\prime}  
+(1-\lambda) y_2^{\prime}$} \\
0 & \mbox{otherwise} .
\end{array}
\right .  
\end{eqnarray}
On taking into account the restriction that no one can have negative money
and setting $y_1^{\prime}+ y_2^{\prime} = L$,
the first integral is proportional to
\begin{equation} 
\int_{y_1}^{\infty} \frac{dL}{(1 - \lambda ) L} 
\int_{\max [0, \{y_1-(1-\lambda)L \} /\lambda]}^{\min [L,y_1/\lambda ]}
dy_1^{\prime} 
f(y_1^{\prime})
f(L- y_1^{\prime}).
\label{Iint}
\end{equation} 
As regards the second integral, at {\it equilibrium} we assume
that the transition 
from $y_1$ to all other levels is proportional to $f(y_1)$.
Also since at equilibrium
$\frac{\partial f(y_1,t)}{\partial t} = 0$, we 
obtain the distribution function to be that
given by Eq. (\ref{distlamb}).

On introducing an upper cutoff $y_c$ for the two party
trading, the contribution 
to the  distribution function $f(y)$
from the microscopic channel becomes
\begin{equation} 
\int_{y}^{\infty} \frac{dL}{(1 - \lambda ) L} 
\int_{\max [0, \{y-(1-\lambda)L \} /\lambda]}^{\min [L,y/\lambda ]}
dx f(x)f(L-x) 
[1- \theta(x-y_c )]
[1- \theta(L-x-y_c )] .
\label{distlambyc}
\end{equation} 

Next, we will analyze the macroscopic contribution to the
distribution function $f(y)$.
The macroscopic contribution to 
$ \left \{ \frac{\partial f(y_1,t)}{\partial t} \right \}_{interaction}$
is given by
\begin{eqnarray}
\left \{ \frac{\partial f(y_1,t)}{\partial t} \right \}_{reservoir ~  interaction}
= &&
\int
r(y_1^{\prime};y_1 )
f(y_1^{\prime},t)
  dy_1^{\prime} 
\nonumber \\
 && 
- \int
r(y_1 ; y_1^{\prime})
f(y_1,t)
  dy_1^{\prime}  .
\label{resinter}
\end{eqnarray}
 Total money involved in trading (between individual and the
macro-system) is $2 y_1^{\prime} (1 - \lambda )$. After interaction, the
resulting money $y_1$ of the individual satisfies the
following constraints
 $\lambda y_1^{\prime} \leq y_1 \leq (2-\lambda) y_1^{\prime}$. 
The first transition rate 
$r(y_1^{\prime};y_1 )$ in the above equation is given by
\begin{eqnarray}
r(y_1^{\prime}, ;y_1) \propto
\left \{ \begin{array}{ll}
 \frac{1}{ 2 y_1^{\prime}
(1-\lambda) } 

& \mbox{if $\lambda y_1^{\prime} \leq y_1 \leq 
(2-\lambda) y_1^{\prime}$} \\
0 & \mbox{otherwise} .
\end{array}
\right .  
\label{resrate}
\end{eqnarray}
As before, at equilibrium 
the second integral on the other hand is proportional to
$f(y_1)$ when only the macroscopic channel is operative.

 On taking into account an upper cutoff $y_c$,
 the contribution to the distribution function
$f(y)$ from the macroscopic channel is
\begin{equation} 
\int_{y/(2-\lambda)}^{y/\lambda} \frac{dx f(x)}{2 x (1 - \lambda ) } 
\theta (x-y_c) .
\label{macy}
\end{equation} 
Then $f(y)$ is given by the sum of the two expressions given
in Eqs. (\ref{distlambyc}) and (\ref{macy})
for the microscopic and the macroscopic channels respectively.
Now it is interesting to note that the solution of the equation
\begin{equation} 
f(y)=\int_{y/(2-\lambda)}^{y/\lambda} \frac{dx f(x)}{2 x (1 - \lambda ) } ,
\label{eqmacy}
\end{equation} 
is given by $ f(y)=c/y^n $.
To obtain $n$ one then has to solve the equation
\begin{equation} 
(2-\lambda)^{n} - \lambda ^ n = 2 n (1 - \lambda ) ,
\label{eqn}
\end{equation} 
and obtains $n=1,2$.
Only $n=2$ is a realistic solution because it
 gives a realistic cumulative probability.
It is of interest to note that the solution 
 is  {\it independent of $\lambda$ } .
Next, we would like to mention that we have chosen 
 the trading probability of a person with certain money,
whether  above the cutoff $y_c$ or below it,
is  determined purely by the number of people 
at that money. However, if the interactions
above $y_c$ occur with a smaller probability governed by 
a prefactor $\alpha$ then Eq. (\ref{eqmacy}) becomes
\begin{equation} 
f(y)=\alpha \int_{y/(2-\lambda)}^{y/\lambda}
 \frac{dx f(x)}{2 x (1 - \lambda ) } .
\label{alphmacy}
\end{equation} 
The solution of the above  equation is again a power law of the
form $1/y^n$ with $n$ depending on {\it both} $\lambda$ and $\alpha$ and is
plotted in Fig. 1.
 When $\alpha < 1$, at smaller values of $\lambda$,
there are two solutions for $n$ with the smaller solution being
less than unity and
 at larger values
of $\lambda$ there is only one solution.
 We have plotted only the larger solution for all values of $\alpha $
as it is the only realistic solution from
a cumulative probability point of view.
 From Fig. 1 we see that 
at a given value of $\alpha < 1 $,
as $\lambda$ increases the value of $n$ and its gradient increases.
Furthermore, for a fixed value of $\lambda $, as $\alpha$ decreases the value
of $n$ increases.

Hence the distribution function, after
simplification, is given by
\begin{eqnarray} 
f(y) = \!\!\!\!\!\!\!\!
&& [1- \theta (y-(2-\lambda ) y_c)]
\left [ \int_{y}^{2y_c} \frac{dL}{(1 - \lambda ) L} 
\int_{\max [0, \{y-(1-\lambda)L \} /\lambda]}^{\min [L,y/\lambda ]}
dx f(x)f(L-x) 
\right .
 \times
\nonumber \\
&&
~~~~~~~~~~~~~~~~~~~~~~~~~~~~~~~~~~~~~~~~~~~~~~~~~~~~
[1- \theta(x-y_c )]
[1- \theta(L-x-y_c )] 
\nonumber \\
&&
~~~~~~~~~~~~~~~~~~~~~~~~~~~~~
+ \left . \alpha  \int_{y_c}^{y/\lambda > y_c} 
\frac{dx f(x)}{2 x (1 - \lambda ) } \right ]
\nonumber \\
&&
+ 
 \theta [y-(2-\lambda ) y_c]
\frac{ f[y-(2-\lambda ) y_c]
 [y-(2-\lambda ) y_c]^n}{y^n} .
\label{distfy}
 \end{eqnarray} 

Here it must be pointed out that Eq. (\ref{distfy}) yields,
 as in the case of the purely microscopic channel without
an upper cutoff, when the savings are zero ($\lambda =0 $) and
$y \rightarrow 0$
 \begin{equation} 
f^{ \prime}(y)
\approx -f(y) f(0) .
\label{distapp}
 \end{equation} 
In obtaining the above equation
we again assumed that the function $f(y)$ and its first
and second derivatives are 
well behaved.
Then the solution for small $y$
is given by
 \begin{equation} 
f(y)
\approx f(0) exp {[-y f(0)]} .
\label{distappsol}
 \end{equation} 

 We solve Eq. (\ref{distfy}) iteratively by choosing a trial function,
substituting it on the RHS and obtaining a new trial function and 
successively substituting the new trial functions over and over again
on the RHS until the difference between the
new trial function $f_n$ and the previous trial function  $f_p$
satisfies the accuracy test $\sum_i |f_n (y_i) -f_p(y_i)| \leq 0.005$.

In Fig. 2 we depict, for the zero savings case ($\lambda = 0 $)
and $\alpha =1$,
$f(y)$ with the cutoff set at ten times the average money per person
and the average money person ($y_{av}$) being set to unity.
 The distribution, as expected, decays exponentially for small values
of $y$ and has power law ($1/y^2$) behavior for $y > 2 y_c = 20$. 
In Fig. 3 we plot $f(y)$ with the cutoff $y_c=0.25$, $y_{av}=1$, and
$\alpha=1$
for values of savings fraction $\lambda = 0.1, 0.5, 0.9$.
Here the power law behavior
($1/y^2$) takes over for $y > (2-\lambda)/4$. At smaller values of $y$
they all become zero with the curves at higher $\lambda$'s
approaching zero faster similar to the case of the purely
two-party trading model (see Ref. \cite{adys}).
Next, in Fig. 4 we show the $f(y)$ behavior at values
of the cutoff $y_c=1/4,1/5,1/6$, $y_{av}=1$, $\lambda = 0.5$,
and $\alpha=1$.
As expected the $1/y^2$ behavior starts at $y=1.5y_c$. 
At smaller values of $y$ the behavior is similar to the purely
two-party trading model studied earlier (see Ref. \cite{adys}).
Lastly, in Fig. 5 we depict $f(y)$ for $y_c =2$, $y_{av} = 1$, $\lambda =0.5$,
and values of $\alpha =0.8, 0.9$. The power law behavior ($1/y^n$)
starts at $y=3$ with $n=4 (3.209)$ for $\alpha=0.8 (0.9)$ \cite{iter}.

In this paper we introduced a new ingredient - interaction of the
rich with wealthy entities and obtain power law behavior similar
to Pareto law. On the other hand the wealth distribution
of the  poorer part of the
society is accounted for through two-party trading.

 The authors are grateful to B. K. Chakrabarti for 
 useful discussions and support.

\begin{flushleft}                                                                 
FIGURE CAPTIONS                                                                 
                                                                                
\vspace {0.5 cm}                                                                
                                                                                
{\bf 1.} Plot of the power law index $n$  versus $\lambda$ 
for values of $\alpha=0.8, 0.9, 0.95, 0.99, 1$.
\vspace {0.3 cm}                                                                

{\bf 2.} Plot of the money distribution function for zero savings
and cutoff being ten times average money.
Both $\alpha$ and the average money per person are set to unity.
\vspace {0.3 cm}                                                                

{\bf 3.} Money distribution function for various savings values
($\lambda = 0.1, 0.5, 0.9$). The average money per person and
$\alpha$ are both set to unity
and the cutoff to one-fourth the average money.
\vspace {0.3 cm}                                                                

{\bf 4.} Money distribution $f(y)$ at $y_{av} =1$, $\lambda =0.5$,
$\alpha=1$,
and various cutoff values ($y_c = 1/4, 1/5, 1/6$).
\vspace {0.3 cm}                                                                

{\bf 5} Money distribution $f(y)$ at $y_c =2$, $y_{av} = 1$, $\lambda =0.5$,
and values of $\alpha =0.8, 0.9$. 
\vspace {0.3 cm}                                                                
\end{flushleft}                                                                 

\begin{thebibliography}{999}
\bibitem{iter}
For convergence, here the iterative procedure for two-party trading
is the same as that employed in Ref. \cite{adys}.
   \vspace {0.2 cm}
\end{thebibliography}
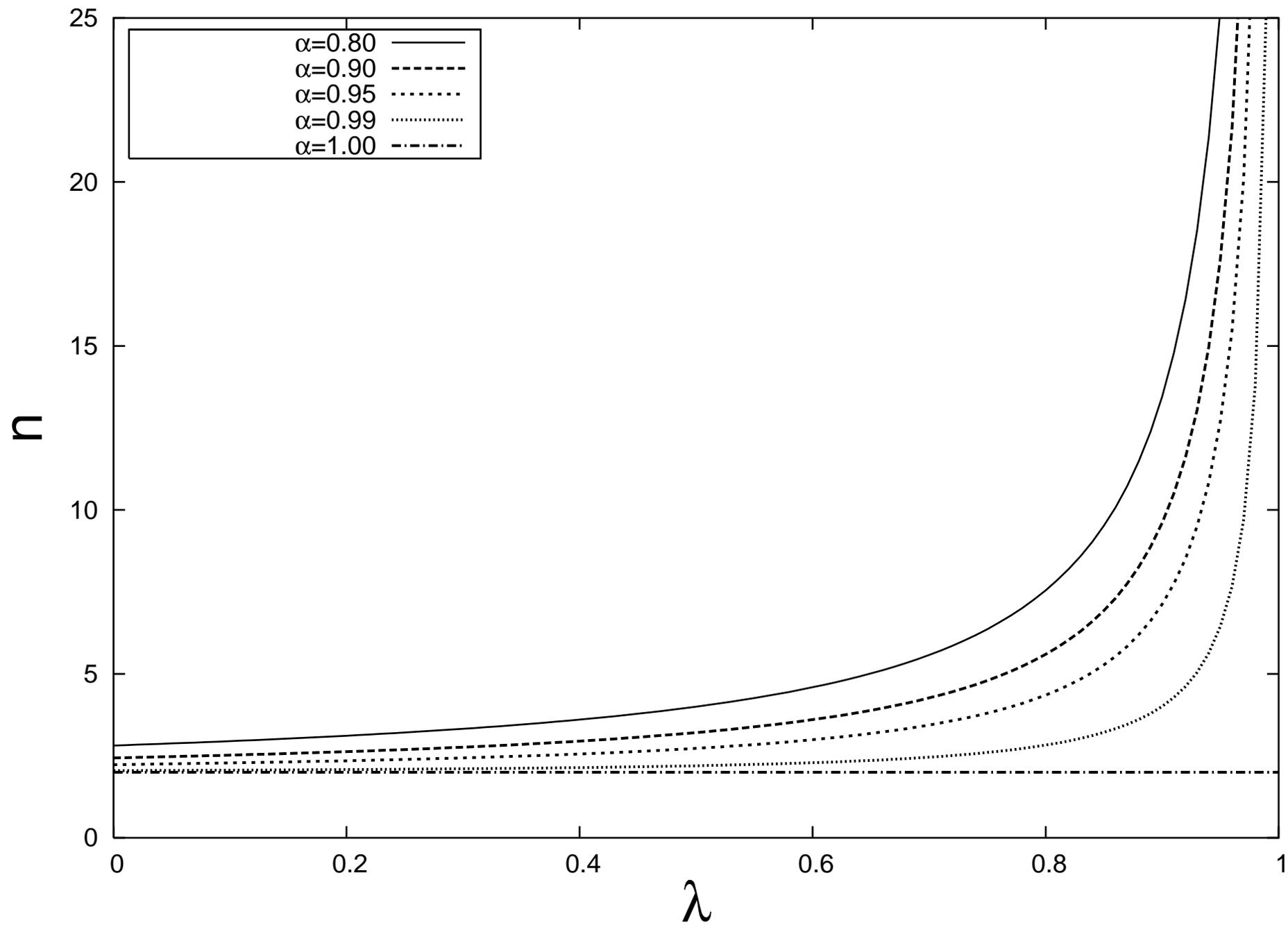
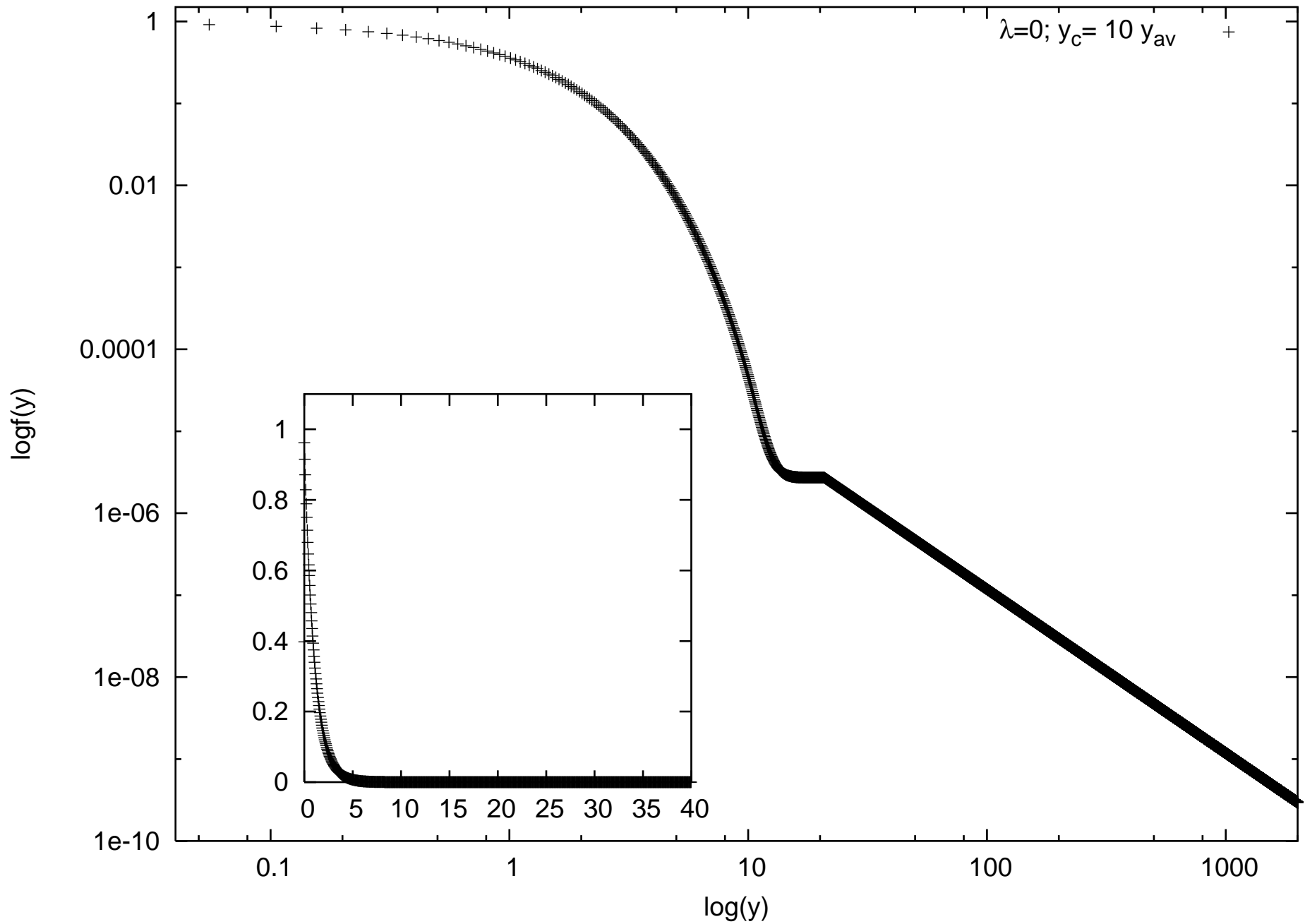
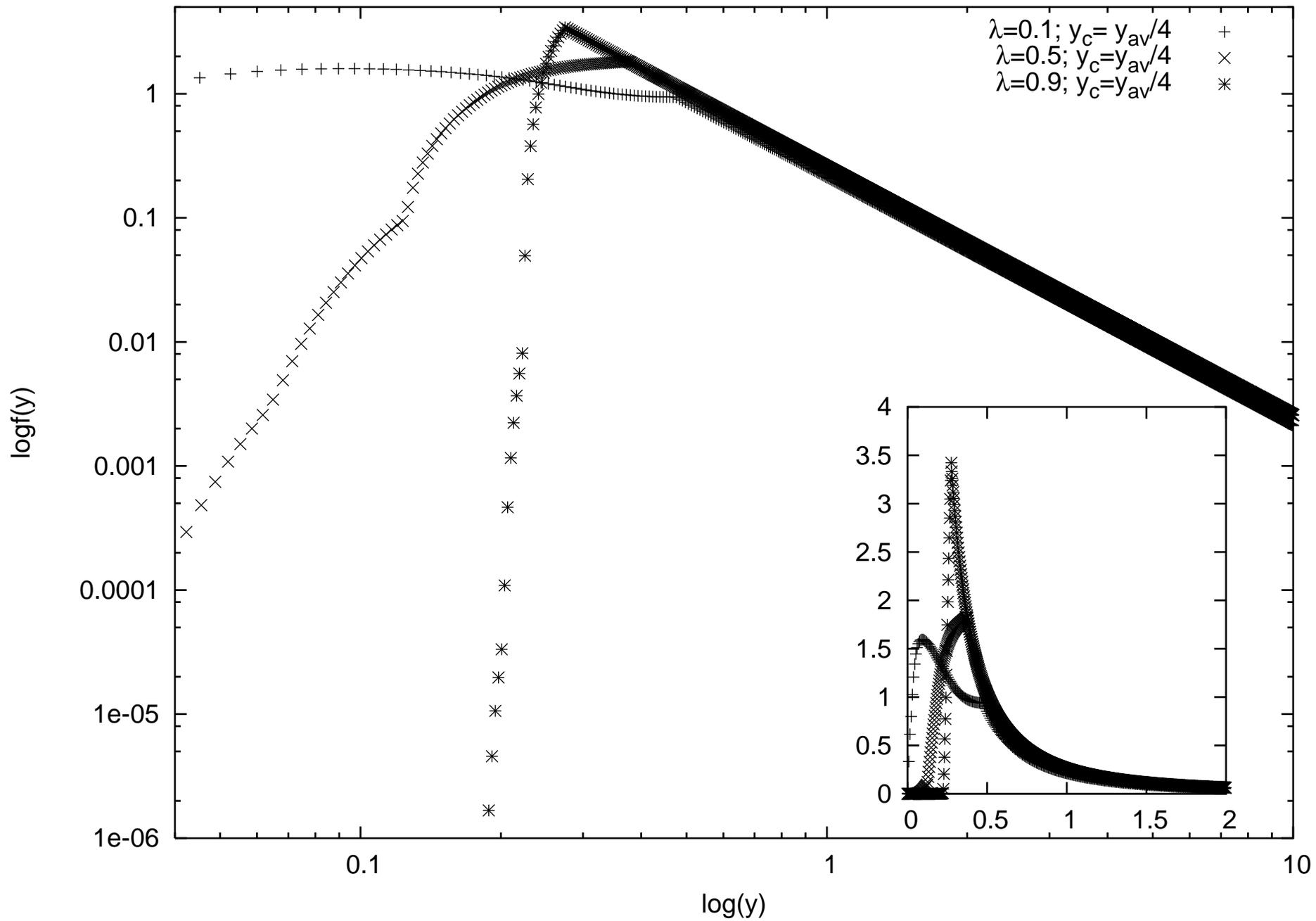
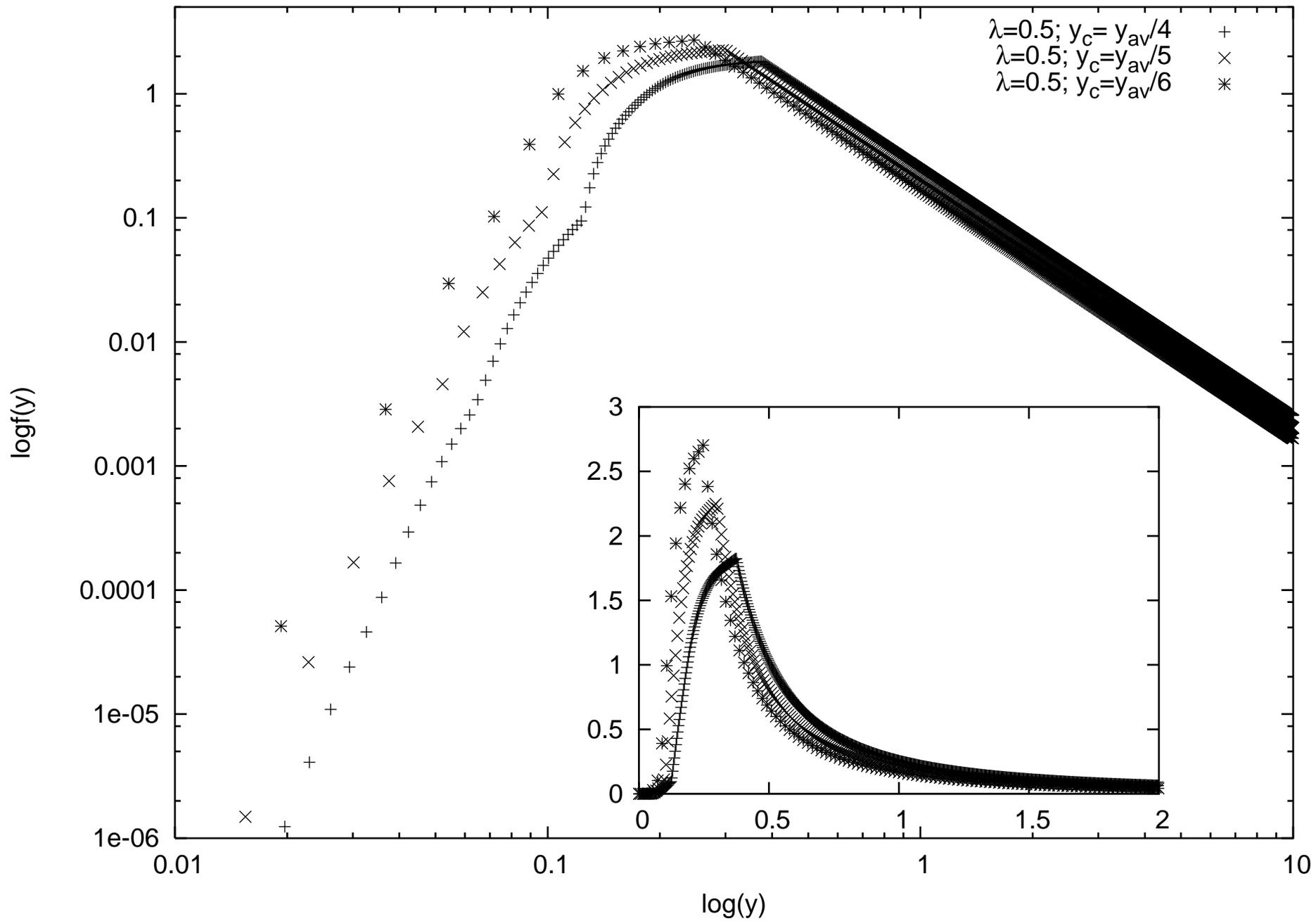
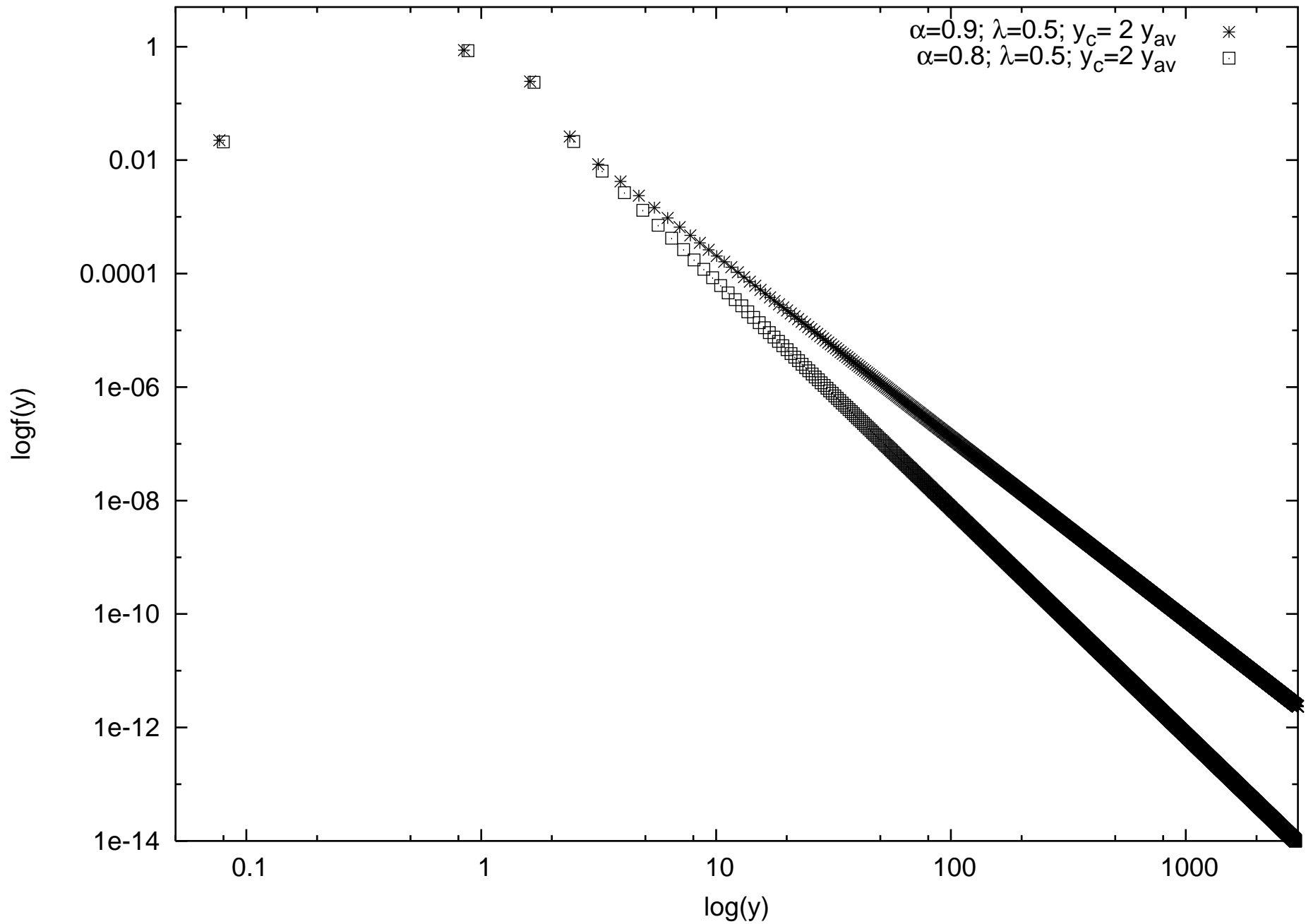
\end{document}